\def\to{\hbox{$\,$--$\,$}}
\def\muspc{\hskip 0.15 em}
\def\mmuspc{\hskip 0.30 em}
\def\mag{\hbox{$\;.\!\!\!^m$}}
\def\etal{\mbox{et~al.}}
\documentstyle{l-aac}
\input psfig
\begin{document}

\thesaurus{13(03.13.2, 03.13.4, 08.08.1, 08.19.1)}

\title{Stellar population synthesis diagnostics}

\author{Yuen K. Ng}

\institute{Padova Astronomical Observatory,
Vicolo dell'Osservatorio 5, I-35122 Padua, Italy}

\date{Received November 4, 1997; accepted March 30, 1998}

\maketitle

\markboth{Yuen K. Ng: Stellar population synthesis diagnostics}{}

\begin{abstract}
A quantitative method is presented to compare 
observed and synthetic colour-magnitude diagrams (CMDs).
The method is based on a $\chi^2$ merit function for a point
$(c_i,m_i)$ in the observed CMD, which has a
corresponding point in the simulated CMD
within $n\sigma(c_i,m_i)$ of the error ellipse.
The $\chi^2$ merit function is then combined 
with the Poisson merit function of the points for which
no corresponding point was found within the $n\sigma(c_i,m_i)$
error ellipse boundary.\hfill\break 
Monte-Carlo simulations are presented to demonstrate 
the diagnostics obtained from the 
combined ($\chi^2$, Poisson) merit function through
variation of different parameters in the stellar population synthesis tool.
The simulations indicate that the merit function can 
potentially be used to reveal information about the initial mass function.
Information about the star formation history of single stellar
aggregates, such as open or globular clusters 
and possibly dwarf galaxies 
with a dominating stellar population, 
might not be reliable if one is dealing with a relatively small 
age range. 
\end{abstract}

\section{Introduction}
In the last decade the simulation of synthetic Hertzsprung-Russell
and colour-magnitude diagrams (hereafter respectively referred to as HRDs
and CMDs) has advanced 
at a rapid pace. It has been applied successfully in various 
studies of young clusters in the Magellanic Clouds (Chiosi \etal\ 1989;
Bertelli \etal\ 1992; Han \etal\ 1996; Mould \etal\ 1997;
Vallenari \etal\ 1990, 1992, and 1994ab),
dwarf galaxies in the Local Group (Aparicio{\muspc\&\muspc}Gallart 1995,
1996; Aparicio \etal\ 1996, 1997{\mmuspc}a,b;
Ferraro \etal\ 1989; Gallart \etal\ 1996a{\to}c; 
Han \etal~1997; Tolstoy~1995, 1996;
Tosi \etal~1991), open clusters in our Galaxy (Aparicio \etal~1990;
Carraro \etal~1993, 1994; Gozzoli \etal~1996),
and the structure of our Galaxy (Bertelli \etal~1994, 1995, 1996;
Ng 1994, 1997{\mmuspc}a,b; Ng{\muspc\&\muspc}Bertelli 
1996{\mmuspc}a,b; Ng \etal~1995, 1996{\mmuspc}a,b, 1997).
\par
Generally all studies focus in the first place on matching the 
morphological structures at different regions in the CMDs
(Gallart 1998). 
In the recent years a good similarity is obtained between the
observed and the simulated CMDs. Unfortunately, the best fit is 
in some cases distinguished by eye. The morphological differences 
are large enough to do this and
the eye is actually guided by a detailed knowledge of stellar 
evolutionary tracks. However, the stellar population technique has 
improved considerably
and an objective evaluation tool is needed, to distinguish quantitatively
one model from another. 
\par
Bertelli \etal~(1992,\muspc1995) and Gallart \etal~(1996c) 
defined ratios to distinguish the 
contribution from different groups of stars. The ratios are defined so
that they are sensitive to the age,
the strength of the star formation burst and/or the slope of the initial
mass function. Vallenari \etal~(1996{\mmuspc}a,b) 
demonstrated that this is a good 
method to map the spatial progression of the star formation in
the Large Magellanic Cloud. Robin \etal~(1996), Han \etal~(1997) and 
Mould \etal~(1997) use a maximum likelihood method to find the best 
model parameter(s), while Chen (1996{\mmuspc}a,b) 
adopted a multivariate analysis
technique. Different models can be quantitatively sampled through
Bayesian inference (Tolstoy 1995; Tolstoy{\muspc\&\muspc}Saha 1996)
or a chi-squared test (Dolphin 1997).
\par
In principle one aims with stellar population synthesis to generate
a CMD which is identical to the observed one. 
The input parameters reveal the evolutionary status of the stellar aggregate
under study. To obtain a good similarity between the observed and
the simulated CMD one needs to implement in the model in the first place
adequately extinction, photometric errors and crowding.
It goes without saying that the synthetic population ought to be comparable
with the age and metallicity (spread) 
of the stellar aggregate. Only with a proper 
choice of these parameters, one can start to study in more 
detail the stellar initial mass function and the star formation
history for an aggregate. 
\par
This paper describes a quantitative evaluation method based on 
the combination of a chi-squared and Poisson merit function%
\footnote{The method introduced in Sect.~2 actually mimics closely
the procedure used in the `fit by eye' method.}. 
It allows one to select the best model
from a series of models. 
In the next section this method is explained and Monte-Carlo 
simulations are made, to display how the diagnostic diagrams 
of the residual points can be employed. 
It is demonstrated that the non-fitting, residual points
provide a hint about the parameter that needs adjustment 
in order to improve the model.  
This paper ends with a discussion
about the method and the diagnostic diagrams together 
with their limitations.
\hfill\break
It is emphasized that this paper deals with a description of 
a quantitative evaluation method for CMDs. Aspects related with
the implementation of an automated CMD fitting program
and comparison with simulated or real data sets will not be 
considered, because they are not 
relevant for the general validity of the method described in the
following section.

\section{Method}
\subsection{The chi-squared merit function}
The method is based on minimizing a chi-squared merit function
between all the observed points in a CMD which have
a corresponding point in the synthetic CMD
within a 3\to5$\sigma(c_i,m_i)$ error ellipse. 
In this range one can
assume that the errors follow a Normal distribution. The $\chi^2$
is a measure for the goodness of the fit 
for $N$\/ observed points in a CMD and is defined as:
\begin{eqnarray}
\quad \chi^2(O,S) & = & \sum_{i=1}^N 
{(c_{i,O}-c_{i,S})^2
\over{\sigma_i^2(c_{i,O})}}
+{(m_{i,O}-m_{i,S})^2
\over{\sigma_i^2(m_{i,O})}} \\
& = & \sum_{i=1}^N 
\left({{\delta m(c_i,m_i)}\over{\sigma_i(c_{i,O},m_{i,O})}}\right)^2 \;,
\end{eqnarray}
where the subscripts ($O,S$) refer respectively to the observed and synthetic
CMD, ($c_i,m_i$) respectively to the colour and magnitude for each 
point $i$\/ in the CMD, 
$\sigma_i(c_{i,O},m_{i,O})$\/ is the error ellipse around each point $i$, 
and $\delta m(c_i,m_i)$ is the 
(colour,{\mmuspc}magnitude) difference between the observed and the 
synthetic star. In addition, the reduced merit function 
$F_\chi$ is defined as:
\begin{equation}
\quad F_\chi = \overline{\chi^2} = \sqrt{\chi^2(O,S)/N_{match}} \;,
\end{equation}
where $N_{match}$ refers to the number of points found within 
3\to5$\sigma(c_i,m_i)$ of the error ellipse for each point between
the observed and the synthetic CMD. Depending on the selection criteria
$\overline{\chi^2}$ is defined to be smaller than or equal to 5. In general
this is one of the functions that should be minimized. Acceptable models
are those with $F_\chi\!\la\!1$, i.e. models for which the difference 
in the (colour,magnitude) of the matched points 
between the observed and synthetic CMDs 
is on average less than 1$\sigma$. 

\subsection{The Poisson merit function}
There are some points which do not have counterparts in the observed
or the synthetic CMD, due to the limits imposed in the comparison. 
For a good fit the number of unmatched points, observed 
and simulated (respectively
$N_{O,not}$ and $N_{S,not}$), should in the ideal case 
be smaller than the Poisson uncertainty 
for the total number of $N_O$\/ observed and  $N_S$\/ synthetic points:
\begin{equation}
\quad N_{O,not} + N_{S,not} \la \sqrt{N_O} + \sqrt{N_S} \;,
\end{equation}
or written as the Poisson merit function $F_P$
\begin{equation}
\quad F_P = {{N_{O,not} + N_{S,not}}\over{\sqrt{N_O+N_S}}} \la 1 \;.
\end{equation}
All the residual points can be placed in a CMD. 
This diagram contains indications 
about parameters that need to be optimized. 
In practice $F_P$\/ will not be smaller than 1,
due to simplifications adopted for the model CMD, to a not optimum
representation of some evolutionary phases or even to
limitations in the transformation from the theoretical
to the observational plane.
In Sect.~3
the diagnostics derived from the CMD 
filled with the residuals are explained in 
more detail. The CMDs filled with the residuals are hereafter also referred
to as the diagnostic diagrams.

\subsection{The global merit function}
Both $F_\chi$\/ and $F_P$\/ span a two-dimensional plane,
see for example Fig.~2, which 
displays the merit for the various models (see Sect.~3). An acceptable model
is obtained when both $F_\chi$\/ and $F_P$\/ are about 1 or smaller.
The best fit can therefore be 
obtained by minimizing the global merit function $F$,
which is defined as
\begin{equation}
\quad F = F_\chi^2 + F_P^2 \;.
\end{equation}
Both $F_P$\/ and $F_\chi$\/ are in units of $\sigma$, say 
$\sigma_P$\/ and $\sigma_\chi$
respectively.
The difference between the observed and synthetic 
points is therefore on average about $\sqrt{F}\sigma$,
where \/$\sigma\!=\!\sigma(F_P,F_\chi)$\/ is a function of 
the average chi-squared difference of
the matched points combined with the
Poisson uncertainty of the unmatched points.
In general the minimization of the global merit function
is mainly due to minimization of the Poisson merit function,
i.e. the reduction of the number of unmatched points. 
An acceptable fit of the data is obtained when $F\!\simeq\!2$ or smaller.
\hfill\break
Note that this procedure is not limited to finding the best
fit for a single-colour, magnitude diagram. It can 
easily be adjusted to fit multi-colour, magnitude diagrams.

\begin{figure}
\centerline{\vbox{\psfig{file=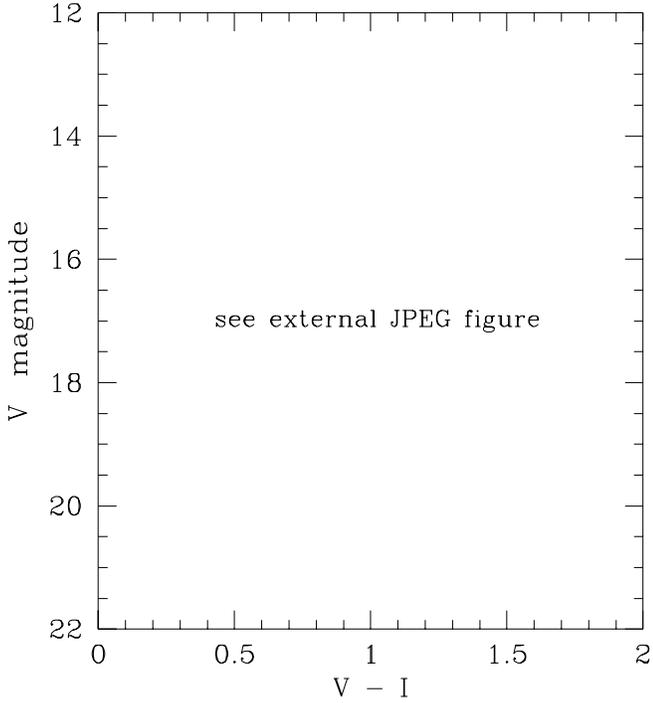,height=9.26cm,width=8.5cm}}}
\caption{The (V,V--I) colour-magnitude diagram of 
the original stellar population
(\mbox{Z\muspc=\muspc0.005\to0.030, t\muspc=\muspc8\to9~Gyr}) 
placed at 8~kpc, see text for additional details.
Note that this simulation reflects a reddening 
free case and that the shape of the red horizontal
branch is due to a large spread in metallicity}
\end{figure}

\subsection{Speeding up}
A comparison between the observed and the synthetic CMD on a point 
by (nearest) point basis can slow down the fitting procedure
considerably, especially when a CMD consists of many data points.
To speed up the whole procedure one can make a concession in accuracy by
binning the $N$\/ data points in $M$\/ average, colour-magnitude boxes 
$(\overline{c}_j,\overline{m}_j)$, each with its own average
error ellipse. Equation (1) can then be re-written to
\begin{eqnarray} 
\chi^2(O,S) & = & \sum_{j=1}^M 
k_j\left(
{(\overline{c}_{j,O}-\overline{c}_{j,S})^2
\over{\overline{\sigma}_j^2(\overline{c}_{j,O})}}
+
{(\overline{m}_{j,O}-\overline{m}_{j,S})^2
\over{\overline{\sigma}_j^2(\overline{m}_{j,O})}}
\right) \\
& = & \sum_{j=1}^M 
k_j\left({{\delta\overline{m}(\overline{c}_j,\overline{m}_j)}
\over{\overline{\sigma}_j(\overline{c}_{j,O},\overline{m}_{j,O})}}\right)^2 \;,
\end{eqnarray}
where $M$\/ is the resolution of the binned CMD, i.e. the number of 
colour-magnitude bins,
and $k_j$\/ is the number of points in each bin.
This method has not been applied here, because 
the gain is small for the low number of objects
used for the examples in Sect.~3.
\begin{figure}
\centerline{\vbox{\psfig{file=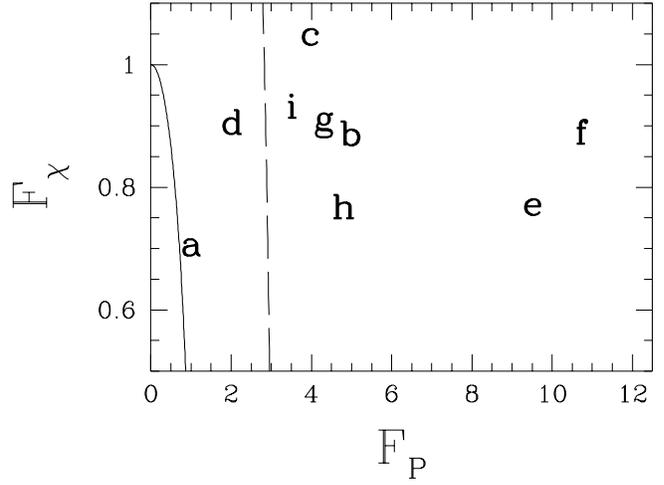,height=6.35cm,width=8.5cm}}}
\caption{The global merit for various simulations. 
The letters {\bf a}\to{bf i} refer to a model for which
the result are given
in Table~1. The solid and long dashed lines denote respectively
the 1$\sigma$ and 3$\sigma$ area, where 
acceptable simulations for the stellar
populations ought to be located. {\tt Model~d} could be acceptable.
However, 
Fig.~3d hints that there is still a systematic discrepancy between
the observations and simulations}
\end{figure}

\subsection{Optimizing with a genetic algorithm}
This paper does not deal with the actual implementation of
an automated search program, which will be the subject of
a forthcoming paper (Ng et al., in preparation).
The following part has been included for completeness 
as an example for a possible approach.
\hfill\break
Genetic algorithms are a class of heuristic search
techniques, capable of finding in a robust way the optimum
setting for a problem (Charbonneau 1995, 
Charbonneau{\muspc\&\muspc}Knapp 1996 and references cited therein).
The optimum setting is searched with a so-called fitness parameter $f$, which
ranges from zero (worst) to one (best).
The fitness parameter $f$\/ can
be expressed as follows in terms of the global 
merit function:
\begin{equation}
\quad f = {1\over{1+F}} \;.
\end{equation}
Acceptable solutions yield $F\!\la\!2$\/ or $f\!\ga\!{1\over3}$. 
An estimate of 
the uncertainty of the input parameters can be obtained by 
doing multiple, time consuming simulations. However, another way 
to estimate the uncertainty is 
to vary for the fittest solution one parameter at a time. 
The fitness $f_{\sigma,k}$
for each parameter $k$ \/ is defined in such a way 
that the global merit $F$\/ changes with 1$\sigma(F_P,F_\chi)$\/ when 
this parameter is varied: 
\begin{equation}
\quad f_{\sigma,k} = {1\over{1+|\sqrt{F_k}-\sqrt{F_{min}}-1|}} \;,
\end{equation}
where ${F_k}$ is the merit for parameter $k$\/ and 
$F_{min}$ is the merit obtained for the fittest population.
This procedure corresponds in Fig.~2 with a jump
in the ($F_P,F_\chi$)-plane from the contour of the optimum solution to 
a contour displaced by exactly 1$\sigma(F_P,F_\chi)$. 
\hfill\break
Note that the contour of the fittest population has the value 
$f_{\sigma,k}\!=\!{1\over2}$ for $F_k\!=\!F_{min}$,
while the contour for the estimated uncertainty has the value 
$f_{\sigma,k}\!=\!1$ for \mbox{$\sqrt{F_k}\!=\!\sqrt{F_{min}}+1$}.

\begin{table*}[t]
\caption{Description of the diagnostic statistics for the 
various models displayed in Figs.~2\to4, see text for 
additional details}
\begin{center}
\begin{tabular}{|l|ccc|cccc|l|}
\hline
model & $N_{O,not}$ & $N_{S,not}$ & $N_{match}$ & $F_\chi$ & $F_P$ 
& $F$ & $f$ & comment \\ 
\hline
\quad a & 73& 70& 4927& 0.70& 1.01& 1.52& 0.398&no variation, see Sect.~3.1\\
\quad b &353&351& 4647& 0.89& 4.98& 25.6& 0.038&distance, see Sect.~3.2\\
\quad c &282&280& 4718& 1.04& 3.97& 16.9& 0.056&extinction, see Sect.~3.3 \\
\quad d &143&143& 4857& 0.90& 2.02& 4.91& 0.169&photometric errors, see Sect.~3.4 \\
\quad e &675&671& 4325& 0.77& 9.52& 91.2& 0.011&crowding, see Sect.~3.5 \\
\quad f &760&759& 4240& 0.89& 10.7& 116.& 0.009&age, see Sect.~3.6 \\
\quad g &306&306& 4694& 0.90& 4.33& 19.5& 0.049&metallicity, see Sect.~3.7 \\
\quad h &340&340& 4660& 0.77& 4.81& 23.7& 0.040&initial mass function, see Sect.~3.8\\ 
\quad i &249&249& 4751& 0.93& 3.52& 13.3& 0.070&star formation rate, see Sect.~3.9\\
\hline
\end{tabular}
\end{center}
\end{table*}

\section{Diagnostics}
Some examples are given with a synthetic population
to elucidate the method and its associated merit function
as described in Sect.~2. It is further shown that
the residual points of the data not matched can provide an 
indication about the parameter to be improved.
A diagnostic diagram is much easier to interpret 
due to the relative small number
of residual points than a CMD with the simulation of the best fit
which may have 5000 or more points.
It might be advantageous for other methods such as ratios, 
maximum likelihood, multivariate analysis or Bayesian inference 
to display the residuals from the best fit obtained.
In this way one can obtain an independent visual impression
of the performance of different methods and even
obtain hints about possible improvements.
\par
The test population used as reference in the simulations has the 
following specifications:
\begin{list}{$-$}{\topsep=0pt\parsep=0pt}
\item a metallicity range, spanning Z\muspc=\muspc0.005\to0.030;
\item an age range from 8\to9 Gyr;
\item an initial mass function with a Salpeter slope;
\item an exponentially decreasing star formation rate with a characteristic time 
scale of 1~Gyr.
\end{list}
This population, displayed in Fig.~1, was found to
contribute significantly to the 
CMD of Baade's Window (Ng \etal~1996a). 
In the simulation the test population is placed at 8~kpc distance. 
The observational limits are \mbox{V$_{lim}$\muspc=\muspc22$^m$} and
\mbox{I$_{lim}$\muspc=\muspc21$^m$}.
In each simulation $N\!=\!5000$ stars are considered.
For simplicity an extinction and crowding free 
case is considered with Gaussian distributed 
photometric errors amounting to $\sigma\!\la$\muspc0\mag05 
per passband. This is then compared with models, in which one of 
the specified parameters is varied. 
For the $\chi^2$-merit function a 3$\sigma$\/ limit 
around each point ($c_i,m_i$) is adopted 
for the range of the error ellipse. 
These variations are then followed 
by comparison with different age, metallicity, initial mass function and 
star formation history.
\hfill\break
Table~1 shows the results of the various simulations 
performed and discussed below, while Fig.~2 displays the
global merit function of these simulations.
Figures~\mbox{3a{\to}i} display the diagnostics diagrams, resulting from the
comparison between the `observed' and synthetic CMDs. 

\subsection{No variation}
The first step is to demonstrate what values one obtains
for $F_\chi,F_P\;{\rm and}\;F$\/ with an acceptable model population.
Such a population 
is obtained from a different realization of the 
test population by modifying the seed of the random number
generator. 
\hfill\break
The values of the merit functions for this simulation ({\tt model~a}) are 
listed in Table~1 and are consistent with the 
expectation that for 5000 points
about 71 points will not be matched in each of the 
observed and simulated dataset.
Figure~2 shows further that this population has 
in the ($F_P,F_\chi$)--plane an average $\sigma$
close to 1.
Figure~3a shows that the residual points 
(open circles for `observations' and crosses for simulations)
are randomly 
distributed over the original (shaded) population.
This is indicative that the {\tt model~a} simulation 
is an acceptable replacement 
for the original population.
\par
Table~2 gives the formal 1$\sigma$ uncertainties determined with eq.~(10).
The estimated error in the distance gives an uncertainty 
in the distance modulus of about 0\mag06. This is a realistic value,
because it is in close agreement
with the 0\mag07\to0\mag08 uncertainty in the distance modulus 
obtained, for example, by Gratton et~al.\ (1997)
for globular clusters.
The uncertainty in the extinction is about 
\mbox{dA$_V\!\simeq$\muspc0\mag06}, which 
is also in good agreement with the best estimates for the
reddening of the globular clusters mentioned above.
They yield \mbox{E(B{\to}V)\muspc=\muspc0\mag02}, which 
is equivalent with an extinction of about 0\mag06
with a standard reddening law.
The \mbox{5\to10\%} uncertainties in the age limits indicate that 
at a 1$\sigma$ level the lower and upper age limit 
are likely not the same, but the whole population might
on the other hand be almost indistinguishable from a population
with a single age of about 8.7~Gyr.
In addition, a 10\% uncertainty is not very different from the 12\% random 
errors estimated for the ages of globular clusters (Gratton et~al.\ 1997).
\hfill\break
\begin{table}
\caption{Formal 1$\sigma$ uncertainties determined with eq.~(10)
for {\tt model~a} (see Table~1 and Sect.~3.1), 
where [Z]\muspc=\muspc{log~Z/Z$_\odot$},
$\alpha$ is the index of the power-law initial mass function 
and $\beta$ is the index of the exponential star formation rate,
specifying its characteristic time scale 
}
\begin{center}
\begin{tabular}{|l|c|}\hline
parameter             & value \\
\hline
log d (pc)            & 3.906 $\pm$ 0.012 \\
A$_V$                 & 0\mag00 $\pm$ 0\mag06 \\
$\log t_{low}$(yr)    & 9.903 $\pm$ 0.043 \\
$\log t_{hgh}$(yr)    & 9.954 $\pm$ 0.023 \\
$[{\rm Z}]_{low}$     & \to0.60 $\pm$ 0.18\phantom{0\ } \\
$[{\rm Z}]_{hgh}$     & +0.18 $\pm$ 0.08\phantom{0\ } \\
$\alpha$              & \phantom{\to}2.35 $\pm$ 0.03\phantom{0\ } \\
$\beta$               & \phantom{\to0}1.0 $\pm$ 1.4\phantom{00\ }  \\
\hline
\end{tabular}
\end{center}
\end{table}
\noindent
The uncertainties in the metallicity range hint that 
an unambiguous determination of the presence of a metallicity range can 
be established. The estimated errors range from 0.1\to0.2~dex,
which is once more a realistic value if one considers
that the uncertainty in the metallicity scale is of the
order of 0.1~dex. Most remarkable is the very small error
obtained for the index of the slope of the power-law IMF.
This is mainly due to the fact that small differences 
in the slope tend to result in large number of 
residuals of main sequence stars for the Poisson merit function.
It strongly suggests that the method described in this paper
potentially could reveal crucial information about the IMF.
This is in high contrast with the estimated uncertainty for the 
index of an exponentially decreasing star formation rate. 
Due to the relative small age range the uncertainty is quite large.
As a consequence, a population with a constant star formation rate 
cannot be discarded.
On the other hand, Dolphin (1997) has
demonstrated that in case of a considerably larger age 
range stronger constraints on the star formation rate could
be obtained. 

\subsection{Distance}
The distance of a stellar aggregate should be determined 
as accurately as possible, because uncertainties in the 
distance modulus can result in a different age for the 
stellar population, see Gratton \etal~(1997) for a detailed 
discussion. For the next simulation ({\tt model~b}) the distance 
of the synthetic population is modified to 8.5~kpc.
Figure~3b displays the diagnostic diagram with the residuals.
\hfill\break
A fraction of the residuals, located
on the main sequence or the sub-giant branch,
are fainter than the original input population (shaded area). 
This feature provides a strong hint that the adopted
population is not located at the proper distance.
It results in a considerably larger value for the 
Poisson merit function. This leads to a rather
large value for the global merit and gives an indication
that the parameters for the matching stellar population 
are not yet optimally tuned. This furthermore appears
in Fig.~2, which shows that the simulation for {\tt model b} 
is not in the 1$\sigma$\to3$\sigma$ range of acceptable solutions.

\subsection{Extinction}
A synthetic population ({\tt model~c}) is made with 
an average extinction of \mbox{$A_V$\muspc=\muspc0\mag10},
to demonstrate the effects of differences in the extinction.
In addition, a random scatter is added to the extinction, 
amounting to 0\mag02.
Figure~3c shows the corresponding diagnostic diagram
with two residual sequences (original versus 
extinction modified), each located
respectively near the blue and red edge of
the original input population (shaded area).
The sequences are also
shifted from each other along the reddening line. 
A comparison between Fig.~3b and 3c further shows that there are
distinct differences between the distribution of the residuals 
when the distance or extinction are not optimally tuned.

\subsection{Photometric errors}
In this simulation the effect of overestimating the photometric
errors ({\tt model~d}) in the simulations is demonstrated. The photometric
errors as a function of magnitude are increased by 50\%.
The results are displayed in Fig.~3d. The figure shows that the residuals
for the modified population are mainly located outside the shaded
region of the original population, while the residuals 
from the original population are almost centered on 
the shaded area.
\par
Figure~2 indicates that model~d might be an acceptable solution,
but Fig.~3d hints that the parameters are not yet optimally tuned.
Figure~3d further displays that the distribution of the residuals
is significantly different from the residuals shown in any of the other panels
of Fig.~3. It is therefore possible to identify  
an inadequate description of the photometric errors 
from the residuals and improve with this information
the description of the simulated errors.

\subsection{Crowding}
In crowded fields single stars can lump together and thus form
an apparent (not always resolved) binary on an image (Ng 1994, 
Ng \etal~1995, Aparicio \etal~1996).
This results in the `disappearance' of some of the fainter 
stars. One can mimic this effect easily in the Monte-Carlo simulations
of a synthetic CMD. One should avoid to include in these 
simulations the completeness 
factors as obtained from artificial star experiments. 
In crowding limited fields one obtains from the simulations 
of apparent binaries the completeness as a function of magnitude.
This apparent, binary induced completeness function should be 
compared with the completeness factors obtained from artificial 
stars experiments.
\par
The crowding simulations ({\tt model~e}) are performed under the assumption
that 5\% of the observed stars in the CMD are (apparent) binaries.
Note that \mbox{$\sim$\muspc10\%} of the total number of simulated stars
are involved in the crowding simulations. 
The actual number is slightly
lower, because the simulations allow for the small possibility
of the formation of multiple star lumps.  
The results are displayed in Fig.~3e. The diagnostics show, except for
a significantly larger scatter, some similarity with those from Fig.~3d.
However, the difference is that one part of the residuals from Fig.~3d  
forms a thinner sequence on top of the lower main sequence band,
while the other part of the residuals is located close to
the lower main sequence band. In the case of crowding the central band
is broader and the remaining residuals are not located near the
lower main sequence band.

\setbox1=\vbox{\vsize5.9cm\hsize5.5cm%
\psfig{file=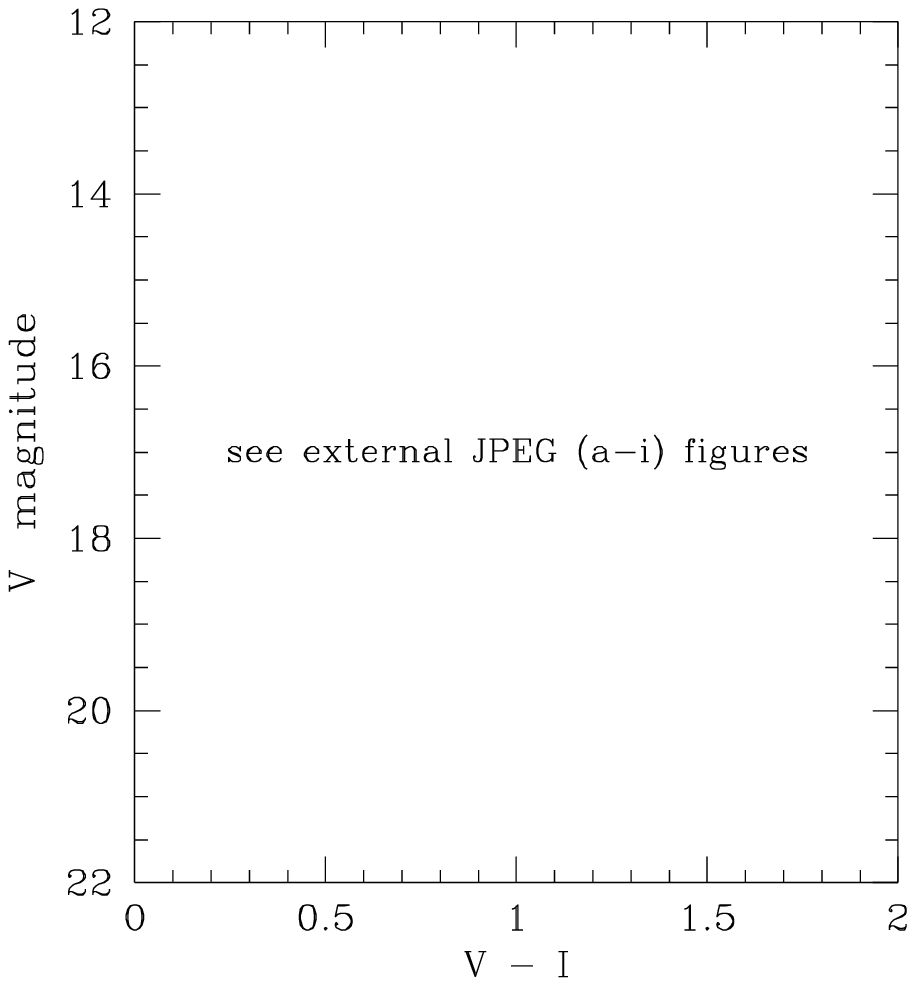,height=6.0cm,width=5.5cm}}
\setbox2=\vbox{\vsize6.0cm\hsize6.2cm%
\psfig{file=fig3jpg.ps,height=6.0cm,width=5.5cm}}
\setbox3=\vbox{\vsize6.0cm\hsize6.2cm%
\psfig{file=fig3jpg.ps,height=6.0cm,width=5.5cm}}
\setbox4=\vbox{\vsize5.9cm\hsize5.5cm%
\psfig{file=fig3jpg.ps,height=6.0cm,width=5.5cm}}
\setbox5=\vbox{\vsize6.0cm\hsize6.2cm%
\psfig{file=fig3jpg.ps,height=6.0cm,width=5.5cm}}
\setbox6=\vbox{\vsize6.0cm\hsize6.2cm%
\psfig{file=fig3jpg.ps,height=6.0cm,width=5.5cm}}
\setbox7=\vbox{\vsize6.0cm\hsize6.2cm%
\psfig{file=fig3jpg.ps,height=6.0cm,width=5.5cm}}
\setbox8=\vbox{\vsize6.0cm\hsize6.2cm%
\psfig{file=fig3jpg.ps,height=6.0cm,width=5.5cm}}
\setbox9=\vbox{\vsize6.0cm\hsize6.2cm%
\psfig{file=fig3jpg.ps,height=6.0cm,width=5.5cm}}
\begin{figure*}
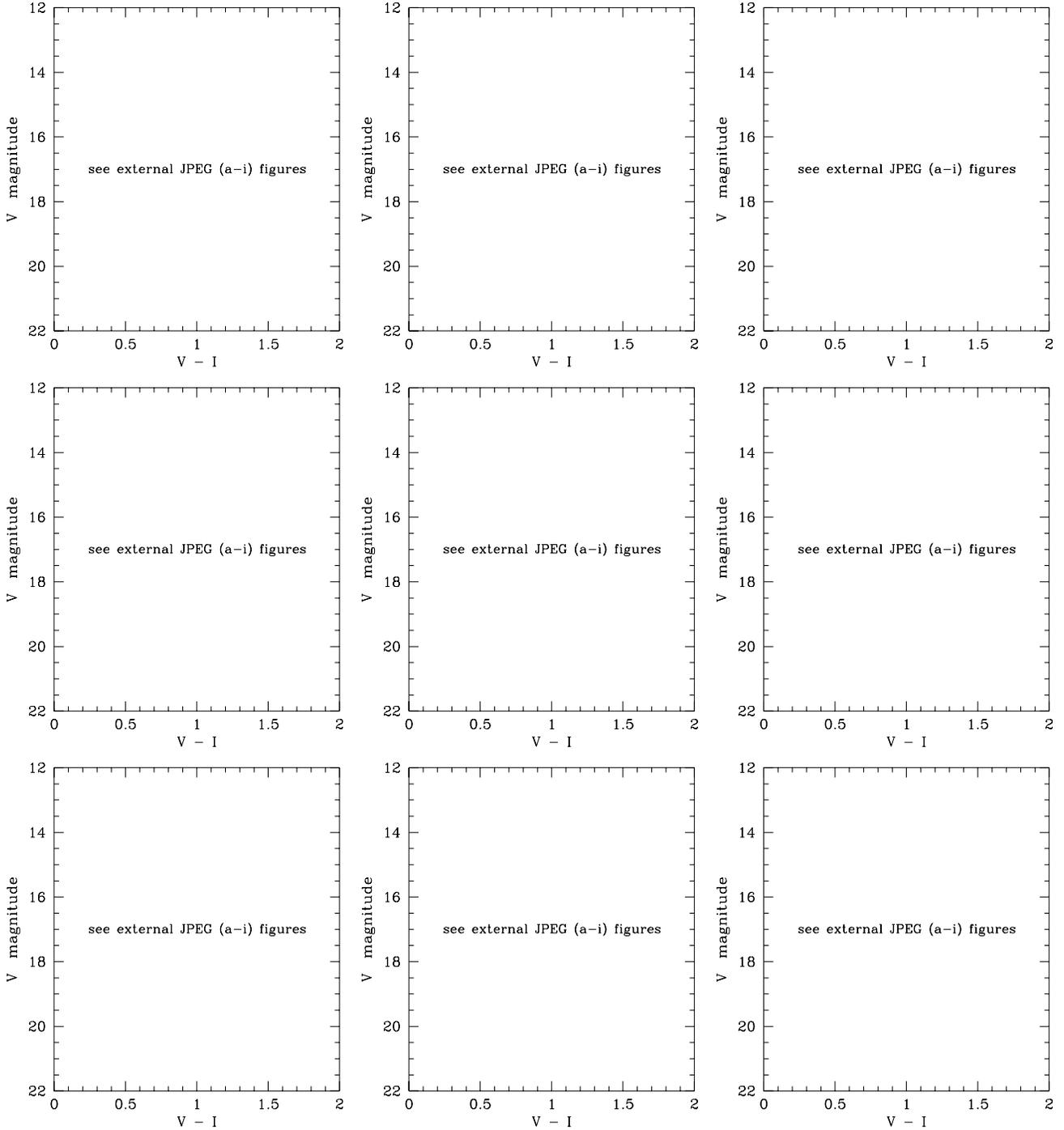

\centerline{\copy1\quad\copy2\quad\copy3}
\medskip
\centerline{\copy4\quad\copy5\quad\copy6}
\medskip
\centerline{\copy7\quad\copy8\quad\copy9}
\caption{Diagnostics diagrams, resulting from the various simulations 
given in Table~1. 
Frame {\bf a}\/ shows the diagram for a different realization,
while non of the major input parameters are varied; 
in frame {\bf b} the distance is varied;
frame {\bf c} shows the diagram when some extinction is added;
frame {\bf d} displays the residuals when different photometric errors 
are adopted;
frame {\bf e} represents the case where 
the amount of crowding is overestimated; 
frame {\bf f} when a younger age is adopted for the stellar population;
frame {\bf g} when the upper metallicity limit is underestimated;
frame {\bf h} shows the effect for a different slope of the 
power-law initial mass function;
and frame {\bf i} shows the case when the star formation rate is
increasing towards younger age, instead of decreasing.
The residual `observed' points which are not fitted are indicated by
open circles, while the residual 
synthetic points are indicated by crosses.
In each frame the shaded area shows approximately the 
part of the CMD covered by the original population. In addition 
the global merit $F$ is reported in each frame}
\end{figure*}

\subsection{Age}
The galactic bulge might contain a rather young stellar population
(Ng \etal~1995, Kiraga \etal~1997), however the results 
by Bertelli \etal~(1995) and Ng \etal~(1996) contradict such
a suggestion. The results by Ng \etal~(1995) might be
induced by the limited metallicity range used in the simulations,
while the results from Kiraga \etal~(1997) do not appear to allow for
a different interpretation. The next simulation ({\tt model f}) is therefore
made with a population which has an age in the range 4\to5~Gyr
and Fig.~3f displays the diagnostic diagram.
At first sight this appears to be similar to the diagram
of Fig.~3c, but there is a pronounced difference: the 
residual sequences are not parallel and furthermore one
of the two sequences is slightly brighter and bluer than
the main sequence turn-off of the original population.  
\par
In Sect.~3.1 it is noted that ages can be determined
to an accuracy of at least 10\%. An analysis of a deep
bulge CMD with the method discussed in this paper might 
well constrain the actual age of the major stellar
populations in the galactic bulge.

\subsection{Metallicity}
In the following simulation ({model~g}) the upper metallicity limit of the
synthetic population is decreased to \mbox{Z\muspc=\muspc0.020}.
Figure~3g displays the diagnostic CMD and shows two almost 
parallel sequences as in Fig.~3c, where the extinction
is varied. The similarity 
is due to the fact that extinction and metallicity differences
show a similar behaviour in (V,V--I) CMDs. Photometry in
other passbands should be explored to avoid this degeneracy.
Ng{\muspc\&\muspc}Bertelli (1996) demonstrate that near-infrared 
(J,J--K) photometry would resolve unambiguously 
the degeneracy between extinction and age-metallicity.

\subsection{Initial mass function}
The sensitivity of the slope of a power-law initial mass function
is demonstrated by changing the slope of the original population
from $\alpha\!=\!2.35$ to 1.35 ({\tt model~h}). Figure~3h displays 
the corresponding diagnostic CMD.
The residuals for the simulation with a shallower slope are dominating
the upper part of the diagnostic diagram, while the residuals from
the original population are concentrated near the faint detection
limit.
The large number of residuals give rise to an increase of the 
value obtained for the Poisson merit function. As mentioned in Sect.~3.1
this behaviour provides a strong constraint in the determination
of the slope of the power-law IMF. Note in addition that the distribution
of the residuals is quite different from the other panels in Fig.~3.
One should realize however that no strong constraint for the power-law
IMF can be obtained when main sequence stars below the turn-off
are not available for analysis.

\subsection{Star formation history}
The final demonstration ({\tt model~i}) invokes a synthetic population
generated for an {\tt increasing} star formation rate
with a characteristic time scale of 1~Gyr.
The uncertainties given in Table~2 (see also Sect.~3.1) already suggest
that a study of the star formation rate cannot be done
reliably when a small age range is considered.
However, the differences in the index of the exponential 
star formation rate are for this simulation suitably chosen,
such that some differences will show up. 
The residuals of the simulation with an increasing star 
formation rate are displayed in Fig.~3i. 
The diagnostic CMD shows two parallel sequences
comparable to the sequence in Figs.~3c and 3g. This 
partly provides an indication of the difficulties involved
in studies of the star formation rate. It further shows once more
that (V,V--I) CMDs are not an optimal choice to study differences
in star formation histories, because it will be difficult
to distinguish differences in the extinction, metallicity,
star formation history and even small age differences
from one another. Additional photometry in other passbands 
ought to be used instead.

\section{Discussion}

\subsection{The RGB and other caveats}
The examples shown thus far are highly idealized cases.
The colours of the stars from a synthetic populations 
are as reliable as the colours of the isochrones. 
Recent analysis of globular clusters (see Reid 1997 
or Gratton \etal 1997
and references cited in those papers) indicate
that a good fit for the main sequence stars
does not necessarily imply a good agreement 
with the stars on the red giant branch. 
The discrepancy partly originates from the uncertainties
in the colour transformations for the late type stars
from the theoretical to the observational plane.
Another cause is likely related to the mixing length 
parameter used in the calculations of the evolutionary tracks.
One therefore has to be cautious in the analysis of 
selected regions in the CMDs with old stellar populations. 
\par
A nice aspect of the diagnostic diagram is that the residuals clearly 
indicate how large the deviations are. In addition,
uncertainties in the treatment of particular evolutionary phases 
might show up in the diagnostics. 
However, this can only be properly evaluated through a massive study, where
one searches for systematic clumping of the residuals in particular 
regions of the CMDs.
The identification of systematic clumps can then be used 
to improve the description of a particular evolutionary 
phase in the computation of new evolutionary tracks.
In contrast to Dolphin (1997) it is argued that one
should avoid the introduction of factors to reduce 
the weight of these particular phases in the fitting procedure.
\par
All these caveats are however not related to the general 
validity of the method presented in Sect.~2. They will
depend on the actual implementation of an automated fitting 
method and they will become important when a comparison with real data 
is made. However, a thorough discussion of problems associated
with the implementation of an automated fitting method 
or a comparison with real data is beyond the scope of this paper.

\begin{figure}[t]
\centerline{\vbox{\psfig{file=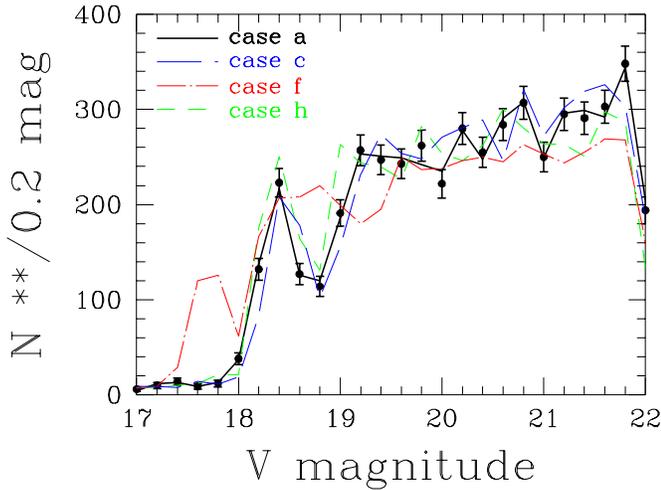,height=6.35cm,width=8.5cm}}}
\caption{Luminosity function for the {\tt cases a, c, f} and {\tt h} (Table~1).
The open circles are obtained from the original input data set. 
The uncertainty per 0\mag2 bin displayed are Poisson error
bars}
\end{figure}

\subsection{Combining the diagnostics}
It will be quite rare that one is going to deal with one of the idealized
diagnostic diagrams from Fig.~3a\to{i}.
More likely the resulting diagnostic diagram is a combination
of these diagrams, indicating that a number of parameters ought to
be modified. One has to remain careful, because
some effects might partly cancel each other out, like for
example age and distance. 
A different distance for the stellar aggregate  
induces a change of the best-fitting age of the stellar 
population (Gratton et al. 1997). 
However, in a CMD the distribution of the stars 
is not exactly the same for populations with a different age.
The subtle differences might not cancel out 
through variation of the distance.
The resulting diagnostic diagram might therefore indicate that yet another 
parameter ought to be optimized - such as the star formation rate -
and in the end indicate that an acceptable fit has been obtained,
while in reality one is dealing with an artifact.
However, one of the major problems in \mbox{(V,V--I)} CMDs remains
the similarity in the behaviour of the extinction,
small age differences, metallicity and the star formation rate.
The results therefore might not always be as reliable as 
they are presented. A heuristic search for the optimum
fit obtained with a genetic algorithm
might properly disentangle the information for these parameters,
but it would be more convenient to avoid this degeneracy 
and use photometry from additional passbands in which this
degeneracy does not occur.

\subsection{Unmatched evolutionary phases}
In general, one will not always find for large amplitude variable stars 
a synthetic counterpart within the error ellipse. Those stars give
rise to a small bias in the Poisson merit function. But, the total
number of large amplitude,
variable stars in any field is expected to be considerably
smaller than $\sqrt{N}$. Therefore, one can ignore in first 
approximation any bias in the results due to variable stars.
\par
Some fast evolutionary phases are not necessarily well described by theory
or even not well covered by the small number of stars observed.
This may lead to the presence of systematic clumps in the CMDs of the
residuals from a massive study. The information obtained from
these clumps can be used to compute new tracks and isochrones.
In many cases however the number of stars present in these 
clumps is expected to be smaller than $\sqrt{N}$. It is therefore
expected that unmatched evolutionary phases in general will not 
affect significantly the search for an optimum fit to the data.
\par
As an aside, Gallart (1998) demonstrates that models are quite
capable to predict subtle details in the observations, despite the
fact that some evolutionary phases are not fully understood.

\subsection{Galactic structure}
In galactic structure studies the stars are distributed along
the line of sight. The diagnostics procedure outlined here is also
useful for these type of studies, in which the observed stars are a complex 
mixture of different stellar populations. Instead of applying a tedious
scheme to deconvolve this mixture in its individual components, it is
more liable to construct a synthetic mixture and compare this directly 
with the observations. The diagnostics will provide in the first place
information about the galactic structure along the line of sight.
Once this has been established, one can explore in more detail 
the initial mass function and the star formation history of the
different stellar populations. It should be possible to obtain some feedback
for the input stellar library with an improved calibration
of the galactic model, and to obtain on the long run 
in a self-consistent way indications 
about the adopted solar abundance partition or enrichment law 
\mbox{$\Delta\,${Y}/$\Delta\,${Z}} (see Chiosi 1996 and 
references cited therein).

\subsection{Open \& globular clusters and the Local Group}
In the studies of open clusters, metal-rich globular cluster
and galaxies with resolved stellar populations from the 
Local Group a considerable amount of fore- and background
stars can be present. It is not easy to take this contribution 
into account, because it is sometimes not clear
if a particular feature is due to stellar evolution
and intrinsic to the aggregate%
\footnote{See for example the suspected intervening stellar population
to the LMC suggested by Zaritsky{\muspc\&\muspc}Lin (1997)
from a feature in the CMD. This interpretation is questioned 
and Beaulieu{\muspc\&\muspc}Sackett (1998) argue
that it is actually intrinsic to the LMC.
Note further that another feature identified as the RGB-bump is most 
likely the AGB-bump (Gallart 1998).}.
One can clean in a statistical sense the galactic 
contribution from a neighbouring field. But this is only possible if
the extinction and photometric errors  
are comparable. Ng \etal~(1996cd) used
a galactic model to account for the contribution of the fore- and background
stars. An unambiguous determination of the age was hampered by the large 
metallicity range and partly by the estimated amount of differential
extinction. In this or other cases the diagnostics scheme as provided 
in this paper might contribute to a significantly deeper analysis.

\subsection{The stellar luminosity function}
Figure~4 displays the luminosity function of the
original population, together with another realization
of this population ({\tt case a}), a population with a modification
in the extinction ({\tt case c}) and age ({\tt case f}), 
and one for which the IMF slope was modified (case h). 
One can easily verify in Fig.~4 that the 
differences between the various populations for the
majority of the magnitude bins are relatively small with respect to the 
generally adopted Poisson error bars.
Only {\tt case f} is significantly different, due to the large age
difference adopted. \hfill\break
For a large number of bins ({\tt models~c} and {\tt h}) 
the number of stars is not exactly within
the 1$\sigma$ uncertainty in case of Poisson errors, but they 
roughly are within 2$\sigma$.
This is a first indication that one did not yet obtain 
an optimum solution. The diagnostic diagrams, Figs.~3c and 3h,
indicate clearly that this is indeed the case.
A comparison with {\tt model~a} further indicates that
the uncertainty in the number of stars
in each bin is slightly over-estimated with Poisson error bars.
This is mainly because the bins are not independent from one another.
An acceptable solution - like {\tt model~a} - should go through
almost every observed point in Fig.~4. About $\sqrt{N_{bins}}$ are expected
to deviate from this expectation. In case of {\tt model~a} about 5~points
might not show a close match in Fig.~4. A close inspection
shows that this is indeed the case. 
\hfill\break
The results indicate that the luminosity functions 
of various realizations -- such as {\tt cases 
a, c}, and {\tt h} -- might apparently not be so significantly different 
from one another and might therefore all be acceptable solutions
for the test population. The values for the
merit function and the diagnostic diagrams (Figs.~3a, 3c and 3h) however 
strongly indicate that only {\tt model~a} is acceptable.
The method presented in Sect.~2, together
with the diagnostic diagram, is more powerful than 
an analysis in which different model luminosity functions 
are compared. 
The global merit function, together with the diagnostic diagrams, gives 
a better discrimination between different models.   
The crucial point lies in the application of the Poisson statistics.
As outlined in Sect.~2,
the only independent quantity to which the Poisson statistics
can be applied is the total number of stars brighter than a 
specific limiting magnitude%
\footnote{For example: the number counts of field SSA~22
(see Fig.~2a from Reid \etal~1997) is within the
Poisson error bars in good agreement with the model 
predictions; their Fig.~2b however shows that too many
stars are predicted by their model, i.e. 61 predicted versus
31 observed for \mbox{I$_{lim}$\muspc=\muspc21\mag5};
this gives $F_P\!>\!5$, indicating that the model does not have
an optimal parameter setting. It is however beyond the objective
of this paper to determine which parameter(s) in their model
has the culprit}.
Suffice to mention that
Table~1 and Figs.~3a, 3c, and 3h show that 
significant differences are present between {\tt cases~c} and 
{\tt h} with respect to {\tt case a}.

\subsection{Future work}
Firstly, an automatic procedure should be developed 
based on the merit functions and the diagnostic diagrams.
A search with a genetic algorithm appears to
be a promising approach. 
As a first test one should apply this program to a synthetic 
dataset, such as the test population used in this paper.
The use of real datasets should be avoided initially,
because unforeseen problems - which are not 
associated with the validity of the method - 
might arise with real data sets.
In particular, problems related with relative fast phases of stellar evolution
or the colour transformation from the theoretical to the observational plane,
see for example Sects.~4.1 and 4.3.
\par
Secondly, a comparison should be made between the results from
an automated search program and the results obtained
from the isochrone fitting technique.
A study of old open clusters in which these techniques 
are applied is underway (Carraro et al., in preparation).
The purpose is to determine if the age of the oldest open cluster
Berkeley~17 (Phelps 1997) is as old as the globular clusters or if 
it has an age comparable to or slightly older than 
the old clusters in the
sample defined by Carraro \etal~(1998).
The next step is to apply this method to the resolved,
multiple stellar populations of dwarf galaxies. However,
with respect to the clusters the results might not be as reliable.
\par
Finally, it is intended to improve through 
\mbox{(self-)} calibration
from studies of essentially single stellar populations,
like open \& globular clusters, the library of stellar
evolutionary tracks and isochrones.

\section{Conclusion}
A method based on the $\chi^2$ merit function is presented
to compare `observed' with synthetic single stellar populations.
Monte-Carlo simulations have been performed to display
the diagnostic power from a CMD containing
the points for which no corresponding synthetic point was
found within a reasonable error ellipse.
The simulations indicate that the CMDs of residual points 
might provide hints about model parameters to be improved.   
The simulations further indicate that one ought to be cautious
with the analysis of stellar luminosity functions and that 
strong hints can be obtained about the 
shape of the initial mass function. 

\acknowledgements{I.~Saviane and L. Portinari 
are thanked for reading of the manuscript.
G.~Bertelli and C.~Chiosi are thanked for their stimulating discussions.
In particular, G.~Bertelli is acknowledged for making available the 
synthetic stellar population generator (HRD-ZVAR),
which produced the stellar populations used for this paper. 
The research of Ng is supported by TMR grant 
ERBFMRX-CT96-0086 from the European Community.}

\end{document}